%% file: arxiv-ndss26.tex
\documentclass[conference]{IEEEtran}
\ifCLASSINFOpdf
\else
\fi
 \usepackage[frozencache,cachedir=.]{minted}

\usepackage{tikz}
\usepackage{amsmath}

\usepackage{booktabs}
\usepackage{listings}
\usepackage{xcolor}
\usepackage{float}
\usepackage{caption}
\newfloat{listing}{tbp}{lop}
\floatname{listing}{Listing}
\usepackage{url}

\usepackage{soul}
\sethlcolor{yellow}
\usepackage{hyperref}

\newcommand{\ourfuzz}{\texttt{\textbf{FivGeeFuzz}}}
\newcommand{\freefivegc}{\texttt{free5GC}}
\newcommand{\ourattack}{Cross-Service Token Attack}

\begin{document}
%
\title{Cross-Service Token: Finding Attacks in 5G Core Networks}


\author{\IEEEauthorblockN{Anqi Chen\textsuperscript{*}}
\IEEEauthorblockA{Northeastern University\\
chen.anqi3@northeastern.edu}
\and
\IEEEauthorblockN{Riccardo Preatoni\textsuperscript{*}}
\IEEEauthorblockA{University of Padova\\
riccardo.preatoni@phd.unipd.it
}
\and
\IEEEauthorblockN{Alessandro Brighente}
\IEEEauthorblockA{University of Padova\\
alessandro.brighente@unipd.it}
\and
\IEEEauthorblockN{Mauro Conti }
\IEEEauthorblockA{University of Padua \& Örebro University\\
conti@math.unipd.it}
\and
\IEEEauthorblockN{Cristina Nita-Rotaru}
\IEEEauthorblockA{Northeastern University\\
c.nitarotaru@northeastern.edu}
\thanks{\textsuperscript{*}Equal contribution.}
}
	

%


\IEEEoverridecommandlockouts

\makeatletter\def\@IEEEpubidpullup{6.5\baselineskip}\makeatother
\IEEEpubid{\parbox{\columnwidth}{
		Network and Distributed System Security (NDSS) Symposium 2026\\
		23 - 27 February 2026 , San Diego, CA, USA\\
		ISBN 979-8-9919276-8-0\\  
		https://dx.doi.org/10.14722/ndss.2026.[23$|$24]xxxx\\
		www.ndss-symposium.org
}
\hspace{\columnsep}\makebox[\columnwidth]{}}

\maketitle

\input{Template/sections/abstract}


%
\IEEEpeerreviewmaketitle

\input{Template/sections/introduction}
\input{Template/sections/background}
\input{Template/sections/Casestudy}

\input{Template/sections/models}
\input{Template/sections/implementation}

\input{Template/sections/evaluation}
\input{Template/sections/relatedWork}
\input{Template/sections/conclusion}





\appendix
\section*{Ethical Considerations}
The work was done on our local computing machines in a containerized environment, so it will not harm any other parties. Our research target is an open-sourced implementation, and we ran our own deployment for testing purposes. We have disclosed our findings to the \freefivegc{} team, and the team members have acknowledged all of them. Patches to fix the problems are under development -- our \ourattack{}, and 6 bugs have already been patched, and the only remaining bug has a patch under development. We also reported our findings to MITRE to request the CVE IDs, and we are still waiting to hear back. 






%
\bibliographystyle{ieeetr}

\input{output.bbl}
\end{document}

%% file: Template/sections/abstract.tex
\begin{abstract}

5G marks a major departure from previous cellular architectures, by transitioning from a monolithic design of the core network to a Service-Based Architecture (SBA) where services are modularized as Network Functions (NFs) which communicate with each other via standard-defined HTTP-based APIs called Service-Based Interfaces (SBIs).  These NFs are deployed in private and public cloud infrastructure, and an access control framework based on OAuth restricts how they communicate with each other and obtain access to resources. Given the increased vulnerabilities of clouds to insiders, it is important
to study the security of the 5G Core services for vulnerabilities
that allow attackers to use compromised NFs to obtain unauthorized access to resources.

We present \ourfuzz, a grammar-based fuzzing framework designed to uncover security flaws in 5G core SBIs. 
\ourfuzz~ automatically derives grammars from 3GPP API specifications to generate malformed, unexpected, or semantically inconsistent inputs, and it integrates automated bug detection with manual validation and root-cause analysis. 
We evaluate our approach on \freefivegc, the only open-source 5G core implementing Release 17-compliant SBIs with an access control mechanism. 
Using \ourfuzz, we discovered 8 previously unknown vulnerabilities in \freefivegc, leading to runtime crashes, improper error handling, and unauthorized access to resources, including a very severe attack we call \ourattack. All bugs were confirmed by the \freefivegc{} team, 7 have already been
patched, and the remaining one has a patch under development.

\end{abstract}


%% file: Template/sections/introduction.tex
\section{Introduction}
5G is the latest standardized mobile network generation, introduced by 3GPP starting from Release 15 \cite{5gsys15, 5gsec15}. 5G popularity is expected to grow from 2.5 billion connections in 2024 to 8.3 billion 5G connections by 2029, representing 59\% of all global wireless technologies \cite{5g_adoption}.
Building on top of 4G and LTE, 5G provides improved service types, higher data rates, more flexibility, and better security. 
5G is the first generation that introduces a dedicated protocol for authentication and key agreement, as well as privacy-enhancing techniques to conceal user identities \cite{5gsys15, 5gsec15, 5gsys18, 5gsec18}.

A critical component of the cellular network architecture is the core network, which manages connections, mobility, security, and services. 
While 4G relied on point-to-point interfaces and network entities for its core design, 5G marks a fundamental change by introducing a Service-Based Architecture (SBA). Specifically, the 5G Core implements its architectural elements as Network Functions (NFs) which interact with other NFs and/or consumer entities via standard-defined HTTP-based APIs, called Service Based Interfaces (SBIs). 
By moving away from proprietary hardware toward cloud-native architectures with containers, microservices, and COTS servers, 
SBA allows operators to reduce both capital and operational expenditures. In addition to proprietary systems, there are several open source 5G Core implementations  \freefivegc \cite{free5gc}, Open5GS \cite{open5gs}, and OpenAirInterface \cite{oai}. These open-source projects are widely adopted in the scientific literature and in commercial products, in particular Open5GS \cite{nextepc} and \freefivegc \cite{fujitsu}\cite{platform9}.

The introduction of the SBA exposes the 5G Core network to new attack vectors as it expands the attack surface (more APIs, more endpoints), it exposes the core to cyberattacks (API abuse and  DoS), and it decreases resilience (a breach in one NF can cascade to others NFs). To address some of these attacks, 
starting from Release 15 \cite{5gsec15, 5gsys15}, 3GPP provides basic security measures for the SBA, defining mechanisms for NF registration, authentication, and access control. For authorization, 3GPP suggests the use of OAuth2 to handle access to resources. 

However, the deployment of the 5G Core in cloud-based infrastructures, especially public clouds 
\footnote{Telefonica Germany \cite{telegerman} and Boost mobile \cite{boostmobile}, migrated several core network services to public clouds such as AWS.}, provides attackers with a novel and previously unavailable access to the core network from inside.
Attackers can now assess the presence of NFs, monitor their traffic, interact with them without the need to access a walled infrastructure, and possibly exploit vulnerabilities to gain privileged access via lateral movements or privilege escalation.
Therefore, although authorization may defend against external attacks, the 5G core may be vulnerable from attacks coming from legitimate/altered NFs.  Given that a significant part of the 5G Core is deployed in clouds and the increased vulnerabilities of clouds to insiders, it is important to study the security of the 5G Core services for vulnerabilities that allow attackers to control some of these services.

Very few works studied the security of the 5G Core. Two recent works \cite{formal5gac,sp25formal5g} used formal methods to analyze the design of the 5G Core's access control mechanisms and identified several vulnerabilities. However, while formal methods provide great benefits in understanding protocol design and identifying design bugs and vulnerabilities, they do not guarantee that deployed services do work as intended.  
Security testing is a complementary approach that 
captures implementation details and deployment conditions 
to identify security issues and software bugs that attackers may exploit such as accessing privileged resources or jeopardizing the network availability. 
To the best of our knowledge, no work has focused on testing all the services in 5G core, whereas the on-going work \cite{syedapifuzzing}\footnote{The code is not publicly available as this is ongoing work.} tested only one interface (out of 6) for only one out of the 10 Core NFs available in \freefivegc.

\textbf{Our contribution.}
In this paper, we focus on testing the entire set of SBI API services of the 5G Core, considering an attacker who compromises an NF or is able to obtain a valid token. 
To demonstrate that such attacks are possible, we first crafted an attack against \freefivegc, the only open source 5G Core implementation supporting OAuth access control. 
The attack, which we call \ourattack, allows an attacker to obtain unauthorized access to resources using a valid token issued for a service, to obtain unauthorized access to resources managed by a different service.


Equipped with proof that such attacks are possible, we set up to systematically evaluate all 5G Core services by using fuzzing. Fuzzing is a testing technique where the tester feeds a program with unexpected, random, or malformed inputs to see if it crashes, misbehaves, or exposes security flaws.
The main design decisions we had to make were:
(1) what is considered correct/incorrect behavior;
(2) how input testing is generated; and 
(3) how does the testing algorithm proceed with respect to previous tests (stateless vs stateful).
We consider correct behavior to be defined 
by the service specifications from the TS 29.x series \cite{release17, 5GC_APIs_Rel17} which defines the SBA APIs and, the SBI specifications and the security aspects of SBA. 
For input testing, a common approach is just to mutate the inputs without any awareness of their meaning. However, the 5G core API specification is complex and defines strict service and interaction models, with pre-defined sets of requests/responses and messages structures. Thus, just using random mutations will not result in finding any meaningful bugs. To address this challenge, we create a detailed grammar for all the services we fuzz. Finally, the service model of the 5G core is stateful, where NFs expect a standard-defined interaction secured by an OAuth2 authorization framework. Our fuzzing algorithm needs to address the statefulness of NFs.


We propose \ourfuzz, a grammar-based fuzzing approach to test 5G Core's SBIs, taking into account their stateful nature.
Our approach builds on the fundamental characteristics of the 5G Core SBA: a standard defined service and interaction model, a stateful service, and the use of an OAuth2 authorization framework.
We design \ourfuzz~ starting from the entire set of 3GPP OpenAPI specifications (48 total) to generate a grammar that describes in details the field of authorization tokens and helps us identify meaningful fields to the protocol design. We use this grammar 
to generate test inputs that are compliant to the expected format, 
and avoid using malformed tokens that would not be accepted.
Finally, leveraging these specifications and their described service model, we can derive testing sequences that account for the stateful nature of NFs' services. As the SBA 5G Core services are implemented using REST API, we leverage RESTler \cite{restler-code}, 
a general stateful REST API fuzzer.

We applied \ourfuzz~ to \freefivegc, an open source 5G core implementation widely adopted as a reference implementation in 5G research and leveraged in industry \cite{fujitsu,platform9}, and the most mature implementation of an access control mechanism, i.e., compliant to 3GPP Release 17 and OAuth. In addition to \ourattack, 
\ourfuzz~ identified 7 previously unknown vulnerabilities in \freefivegc: 5 in the Unified Data Management, one in the Network Slice Selection Function, and one in the Policy Control Function.
We responsibly disclosed them to \freefivegc{} team: they acknowledged our findings and are developing corresponding patches---with 7 patches already done. 
We summarize our contributions:
\begin{itemize}

    \item We crafted \ourattack, a severe access control verification implementation error in \freefivegc{} that allows an attacker to obtain a token for a service and then use it for another service. 
    
    \item We designed \ourfuzz, a grammar-based fuzzer to test the direct communications among 5G Core network SBIs for unhandled error messages, misconfigurations, or authorization failures.
    
    \item 
    We built the grammar for \ourfuzz{} using the entire
    set of OpenAPI specifications (48),
    and tested all 10 services implemented in \freefivegc, identifying 7 previously unknown bugs:
    6 bugs cause a runtime panic to the UDM, NSSF, and PCF services, while 1 bug causes incorrect error handling in the UDM. \ourfuzz~ also automatically rediscovered our \ourattack.
    All bugs were confirmed by the \freefivegc{} team, our \ourattack{}, and 6 bugs have already been patched, and the only remaining bug has a patch under development.
    
\end{itemize}



%% file: Template/sections/background.tex
\section{Background}

In this section, we provide a brief overview of the 5G Core architecture, its service-based interface (SBI) and the access control mechanism, and an implementation  of 5G Core 
with access control mechanism-- the \freefivegc~ platform. We then discuss security goals and previous vulnerabilities against 5G Core and describe the focus of our work.

\subsection{5G Core Network Overview}


The 5G service architecture consists of three main components: the User Equipment (UE) or end-user devices, the base station Radio Access Network (RAN), and the core 5G network functions (5G Core or 5GC).  UEs are end-user devices, such as smartphones with a SIM card provided by the network operator, and it can communicate with RAN to get access to the core network. The 5G RANs, also named Next Generation Node B (gNBs), are the base stations located in many physical places to help UEs communicate with the core network. The 5G Core is responsible for services including connection, mobility management, user authentication, and subscriber data handling. It also supports two modes of communication: non-roaming mode, in which the user connects to services offered by their own network operator, and roaming mode, in which the user can continue to access services while switching between different network providers. 

\textbf{Service-Based Architecture (SBA).}  
In 5G, the core network shifts from a hardware-based to a Service-Based Architecture (SBA), where the functionality is partitioned into multiple Network Functions (NFs), each responsible for a specific set of tasks. Numerous NFs compose the 5G SBA, including the Network Repository Function (NRF), which maintains a registry containing NF instance profiles and their available services, the Unified Data Management (UDM), responsible for managing subscriber data, the Access and Mobility Management Function (AMF), which handles user equipment registration and mobility, the Policy Control Function (PCF), which enforces policy rules, and the Network Slice Selection Function (NSSF), which assists in assigning user sessions to the appropriate network slice.

\begin{figure}[htbp]
    \centering
    \includegraphics[width=0.95\linewidth]{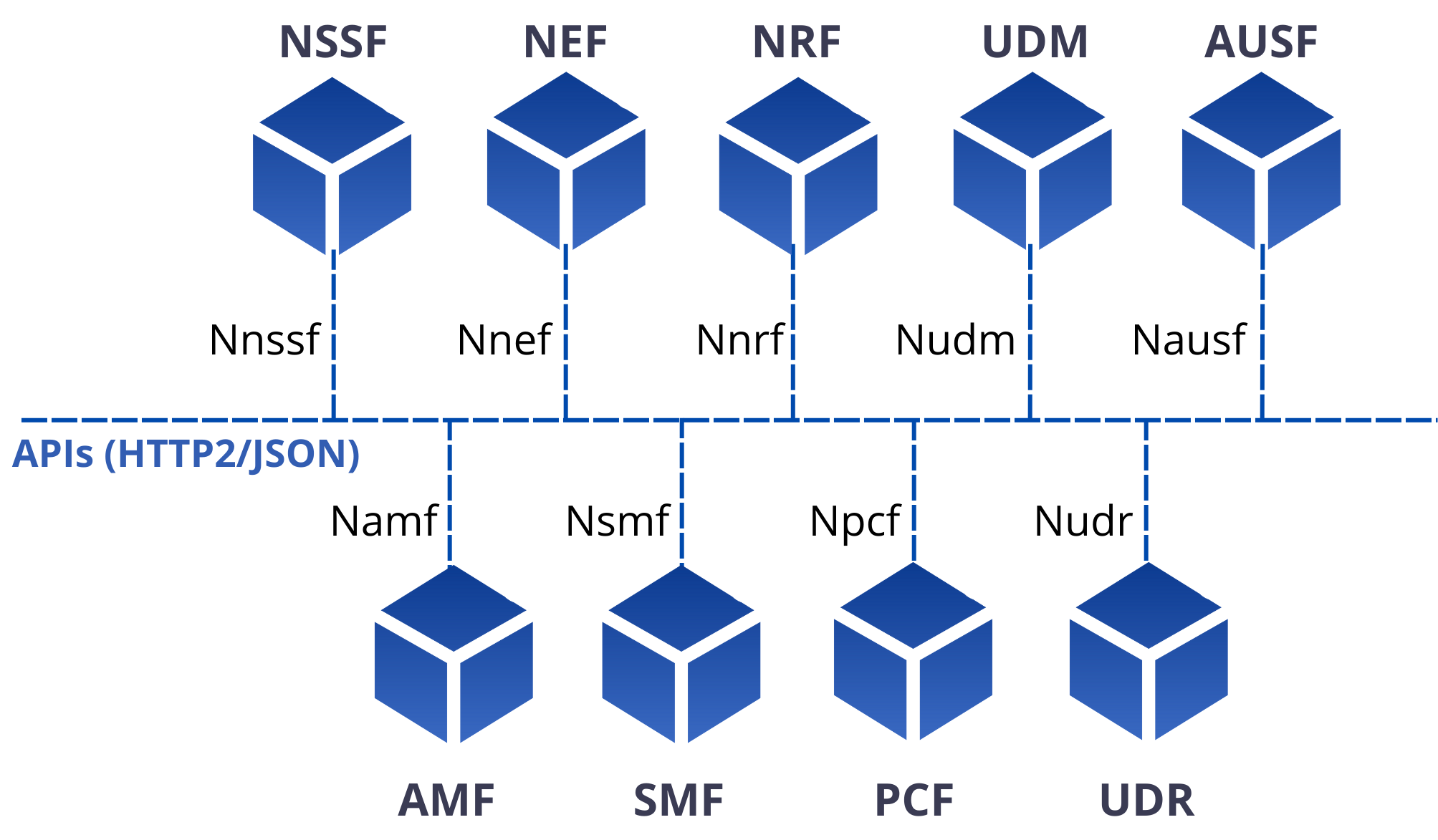}
    \caption{Overview of main 5G Core SBIs and related NFs. 
    }
    \label{fig:SBI}
\end{figure}

NFs interact with each other by exposing a set of capabilities as services and communicating through Service-Based Interfaces (SBIs). The SBI is a framework of RESTful APIs standardized by the 3GPP, which defines multiple interfaces such as \textit{Nudm}, \textit{Nnrf}, and \textit{Nsmf} (Fig. \ref{fig:SBI}). Each interface groups one or more related services, defined according to the OpenAPI v3 specification\cite{5GC_APIs_Rel17}, and communicates using JSON over HTTP/2 at the application layer. 

\textbf{NF Communication Model.}
Each NF can act either as a service producer, by exposing one or more APIs that implement specific capabilities, or as a service consumer, by invoking the services offered by other NFs. For example, the \textit{Nudm} interface connects consumer NFs to the UDM, and includes services such as \textit{Nudm\_SDM} for subscription data management and \textit{Nudm\_UEAU} for user authentication. Also, each NF service can group many NF operations. For example, AMF provides an NF service \textit{Namf\_Communication}, under which there exist several operations like \textit{UEContextTransfer}, \textit{ReleaseUEContext} \cite{5gproc18}, etc. Both service-level scopes and operation-level scopes can be specified in a service request and response. NFs can communicate directly with each other, or indirectly through a Service Communication Proxy (SCP), which acts as a middleware/broker. 

\textbf{Network Slicing.}
5G supports network slicing, reusing the same physical network architecture (the RAN, transport network, and core network) by generating different logically-separated virtual network slices, to facilitate customized service requirements, such as bandwidth, latency, and even security capabilities and characteristics. A Network Slice can be identified by Single Network Slice Selection Assistance Information (SNSSAI). 
A Network Slice instance includes a set of NF instances and the required resources (e.g., compute, storage, and networking) and is identified by Network Slice Instance Identifier (NSI ID) \cite{5gsys18}. 
The NRF can orchestrate NF access control by specifying the Network Slice scopes, enabling fine-grained control over service exposure and authorization across logically isolated slices.

\subsection{5G Core Access Control}
\label{subsec:5gac}
While the SBA improves flexibility, scalability, and interoperability across vendors, it also increases the exposure of critical control plane functions to potential misuse. Not all NFs should be able to access all services; only authorized NFs should invoke services they are allowed to.

\begin{figure}[htbp]
    \centering
    \includegraphics[width=0.95\linewidth]{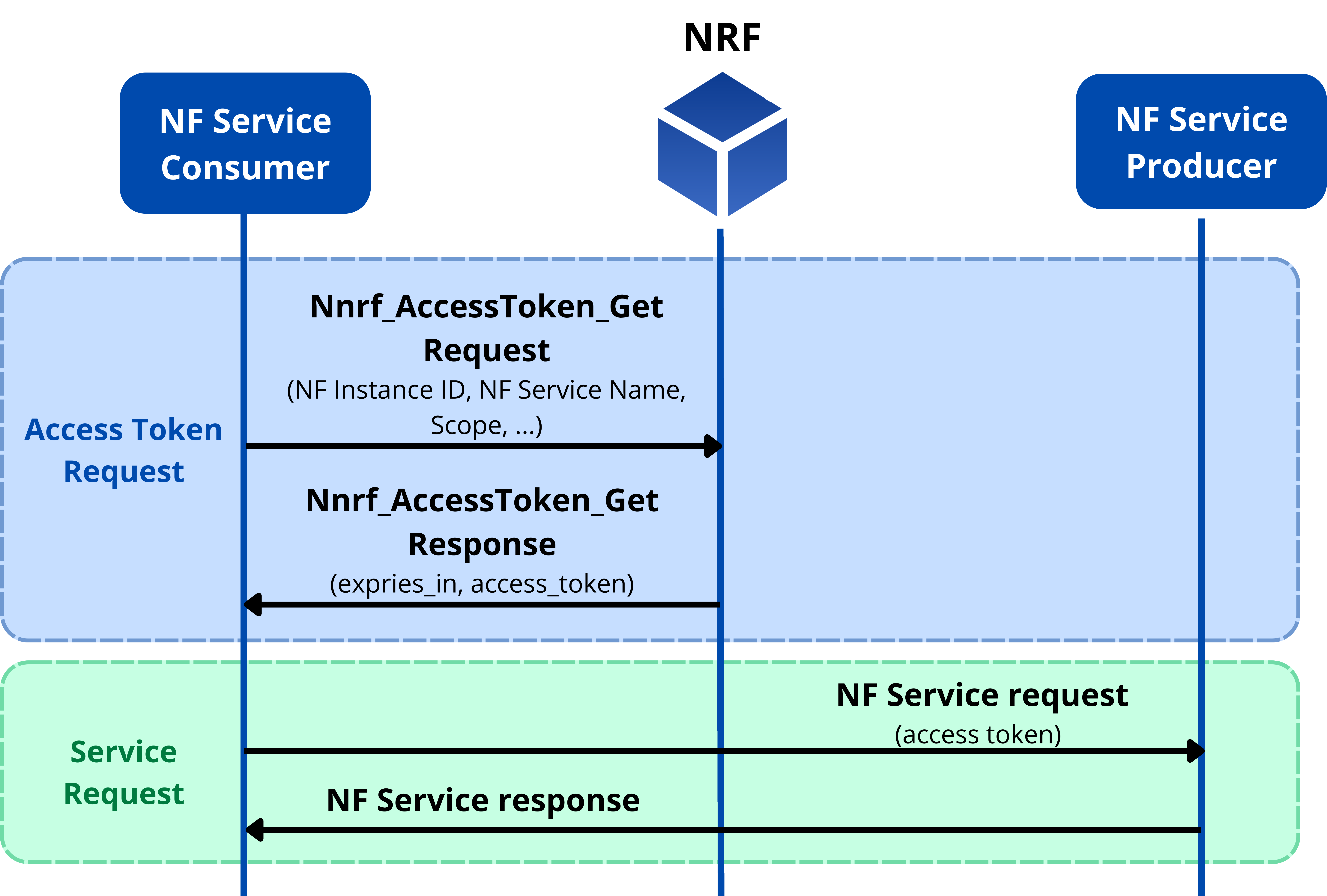}
    \caption{Two-step authorization workflow in the 5G Core. 
    }
    \label{fig:auth_flow}
\end{figure}

To provide this access control and authorization mechanism, 3GPP selected the OAuth 2.0 protocol \cite{rfc6749} as the standard for authorizing access to SBI services \cite{5gsec15}. OAuth 2.0 supports several grant types for authorization purposes, and 5G Core mandates the use of the Client Credentials Grant. The OAuth 2.0 roles are mapped as follows \cite{5gsec15, 5gsec18}:
the NF Service Producer acts as the OAuth 2.0 resource server,
the NF Service Consumer acts as the OAuth 2.0 client,
and the NRF acts as the OAuth 2.0 authorization server. 
When OAuth2.0 configuration is enabled, an NF Consumer requesting a service from an NF Producer needs to get an authorization token from the NRF. The access token should include a scope reflecting the request. Only with a valid matched access token, the NF Consumer can obtain access to the target services (Fig. \ref{fig:auth_flow}). 

\textbf{Prerequisite Steps.}
Before proceeding with the request and issue of the access token, there are some prerequisite steps. First, the NF Service Consumer should be registered with the NRF through the \texttt{Nnrf\_NFManagement\_NFRegister} Request, using its NFInstanceID as the client ID during registration. The NF Service Producer should also be registered with the NRF, and the producer's NF Profile may indicate the resources and the service operations that are allowed per NF type of the NF Service Consumer or per NFInstanceID of the consumer. Meanwhile, the NRF and NFs should have mutually authenticated each other, where the NFs are identified by the NFInstanceID recorded in the public key certificate. The NRF and NF should also share the required credentials (e.g. the NRF's public key or a shared secret)\cite{5gsec15, 5gsec18}.



\begin{table*}[ht]
\centering
\caption{Comparison of open-source 5GC implementations}
\begin{tabular}{p{2.2cm}|p{8.8cm}|p{2.2cm}|p{2.3cm}}

\toprule
Platform & Components & Specification  & Access Control\\ 
\midrule
 \freefivegc \cite{free5gc} & \textbf{10 Core NFs} (AMF, AUSF, CHF, NEF, NRF, NSSF, PCF, SMF, UDM, UDR), N3IWF, TNGF, and UPF \cite{free5gccode} & Release 17 \cite{5GC_APIs_Rel17} & Yes \\
 Open5GS \cite{open5gs} & \textbf{8 Core NFs} (AMF, AUSF, NRF, NSSF,
PCF, SMF, UDM, UDR), SCP, SEPP, BSF, and UPF \cite{open5gsdoc} & Release 17 \cite{5GC_APIs_Rel17} & No \\
 OAI \cite{oai} & \textbf{7 Core NFs} (AMF, AUSF, NRF, NSSF, SMF, UDM, UDR) and UPF \cite{oaicode} & Release 15 \cite{5gsys15} & No 
 \\
\bottomrule
\end{tabular}

\label{tab:open5gc}
\end{table*}

\textbf{Access Token Request.}
Once the prerequisite conditions described above are satisfied,  the NF Service Consumer requests an access token from the NRF via the \texttt{Nnrf\_AccessToken\_Get} operation (Fig. \ref{fig:auth_flow}), providing its NFInstanceID, the target NF type or NFInstanceID, and the target service scope, which specifies the exact service or set of services the requester seeks to access. The NF Service Consumer may also include a list of SNSSAIs or a list of NSI IDs for the expected NF Service Producer instances in the access token request.

\textbf{Access Token Issued.}
Upon receiving the request, the NRF checks whether the NF Service Consumer is authorized to access the requested services by comparing the request parameters against the relevant authorization attributes of the target NF Service Producer, such as its allowed NF types and permitted network slices that are known from the NF registration phase. If the request is authorized, the NRF generates a signed access token and returns it to the consumer (Fig. \ref{fig:auth_flow}). 

In the 5G Core, the access token, implemented as a JSON Web Token (JWT) \cite{rfc7519}, includes a set of items such as the issuer (the NFInstanceID of the NRF), the subject NF, the audience NF, the scopes, and the valid period. The integrity of the access token is secured by either a digital signature signed with NRF's private key or Message Authentication Codes (MAC) generated with a shared secret \cite{5gsec15, 5gsec18} based on JSON Web Signature (JWS) \cite{rfc7515}. 

\textbf{Access Token Validation and Usage.}
In the second step, the NF Service Consumer includes the access token in a service request to the NF Service Producer (Fig. \ref{fig:auth_flow}). The producer first verifies the token’s integrity by checking the signature with either the NRF's public key or a shared secret. If a list of SNSSAIs or a list of NSI IDs exists for the producer NF type, the NF Service Producer needs to check that at least one of the SNSSAIs or NSI IDs served by itself is included in the list. Then the producer checks that the audience matches its NF type or NFInstanceID, confirms that the requested operation is permitted, and ensures that the token has not expired \cite{5gsec18}. If all checks pass, the producer can proceed to provide the requested service (Fig. \ref{fig:auth_flow}).

Authorization granularity is governed by the scope claim in the token. Service-level scopes authorize access to specific services for an NF type, while operation-level scopes provide finer control by restricting access to individual operations or data subsets within a service. These additional scopes are optional and depend on the NF Service Producer’s configuration, enabling more precise permission management and reducing the risk of over-privileged access in the 5G Core. 

\subsection{Open-Source Implementations of Access Control in 5G Core}

Currently, there are three open-source implementations of the 5G Core, each with different levels of functionality (see Table \ref{tab:open5gc}). 
OpenAirInterface (OAI) \cite{oai} is an open-source software project now maintained by the OpenAirInterface Software Alliance (OSA). \freefivegc{} \cite{free5gc} has become a Linux Foundation project in 2024. Finally, Open5GS is a community project, also commercialized by NewPlane \cite{newplane}.
\freefivegc{}\cite{free5gc} and Open5GS \cite{open5gs} are compliant with 3GPP Release 17 \cite{5GC_APIs_Rel17}, while OAI \cite{oai} is only compliant with 3GPP Release 15 \cite{5gsec15, 5gsys15}. 
\freefivegc~ has implemented the largest number of core NFs (10), 2 more than Open5GS and 3 more than OAI. 
 
\freefivegc~ is the only publicly available 5G Core implementation that includes a Service-Based Interface (SBI) implementation compliant with 3GPP Release 17 supporting OAuth 2.0 as its access control framework (Table \ref{tab:open5gc}).  
This makes it a suitable candidate for evaluating both robustness and access control enforcement in state-of-the-art SBI deployments.  While \freefivegc{} is the most mature implementation, it still lacks some features that are specified in the 3GPP specifications of release 17 \cite{5GC_APIs_Rel17} and lacks support for more sophisticated features such as indirect communication where NFs communicate through a proxy (SCP),  and roaming where the visitor NRF and home NRF need to coordinate.

\subsection{Security of 5G Core}

The main security goals for 5G follow typical
security properties such as confidentiality, integrity,
authentication, access control, availability, and anonymity.
Some of these properties apply to all components of the 5G architecture, while others are specific to some components. For example, integrity applies to all communication involving the UE, RAN, and Core, while privacy applies to the UE when users are roaming through other networks than the home network.

As the focus of this work is the 5G core network, below we describe how an attacker can target the security goals of the core network by leveraging the large attack surfaces for 5G represented by UE, the RAN, and the Core itself, e.g., the SBA architecture.  We provide a summary in Table \ref{tab:bksecurity}.

\textbf{Security goals of the 5G core}.
One of the main goals of the core is to protect the SBA architecture. While the communication takes place over
HTTP/2, TLS is not mandatory \cite{tls5g}, thus communication can take
place unencrypted, allowing attackers with access to the
cloud where SBA is deployed to monitor, intercept, and change the communication.

From the perspective of core network access control, an NF should only be allowed to access the resources that it is authorized to, and never gain any unauthorized access to other NF services that might cause a sensitive data breach or further attacks due to privilege escalation. 

Further, the core network interactions should be robust against any unexpected messages from an attacker who compromised an NF or pretended its identity to interfere with the inter-NF communication.  

\textbf{Previous attacks.}
Previous work has used both testing and formal analysis to study attacks against the 5G core conducted through UE or the SBA architecture.  Specifically, previous work has shown how an attacker can send unexpected malformed messages from a UE via RAN to the core network to disrupt the service, achieve authentication bypass, or conduct billing fraud \cite{ransacked, corecrisis, 5gfuzzinfocomm, amfuzz}. 

Very few works studied attacks conducted directly against the core.
The work in \cite{formal5gac, sp25formal5g} used formal methods to analyze the design of the 5G Core itself, and found several attacks, including unauthorized access to sensitive information, illegitimate access to services, and denial of service.
With respect to finding attacks in 5G Core implementations, the only work we are aware of is the on-going work \cite{syedapifuzzing}, which focused on limited testing of the robustness of the Event Exposure API of the AMF in \freefivegc{} and found 2 vulnerabilities. \freefivegc{} currently supports a total of 10 NFs that have not been assessed for vulnerabilities.

\begin{table*}[t]
\centering
\caption{Summary of Previous Attacks against 5G}
\begin{tabular}{p{2cm}| p{2.5cm}|p{6cm}|p{2.3cm}|p{2.0cm}}
\toprule
Attack surface & Attacker Goal & Attacker capabilities & Design Analysis & Code Testing \\ 
\midrule
UE  & Disrupt the core network  & Control the compromised UEs with operator-provided SIM to send arbitrary messages via benign RAN to 5GC &   & \cite{corecrisis} \\
UE and RAN & Disrupt the core network &  Modify, inject, replay, and drop packets between the access network and the core network via compromised RAN I/O &   & \cite{5gfuzzinfocomm} \\
Core NFs (direct and non-roaming mode) & Access priviledge escalation (resources or operations) & Full control with a compromised registered consumer NF &  \cite{formal5gac} &  \\
SCP and Core NFs (indirect /roaming mode) & Access priviledge escalation (resources or operations) &  A compromised Service Communication Proxy(SCP), a compromised NF, and malicious visting NF & \cite{sp25formal5g} &  \\

Core NFs (direct and non-roaming mode) & Access privilege escalation or disrupt the core network &  A compromised registered consumer NF or the attacker can send messages in the no-TLS-protected core network  &  & \ourfuzz \\
\bottomrule
\end{tabular}

\label{tab:bksecurity}
\end{table*}

\subsection{Our focus}
In our work, we focus on \freefivegc{} \cite{free5gc} as a core network implementation.
\freefivegc{}, hosted by the Linux Foundation as open-source code, is widely adopted \cite{syedapifuzzing, corecrisis, formal5gac, sp25formal5g}. 
Our work is conducted on \freefivegc{} v4.0.0 (the newest version at the paper writing time is v4.0.1 recently released on Apr. 22nd 2025), which is the most mature open-source 5GC implementation, conforming to a recent specification, 3GPP Release 17 \cite{5GC_APIs_Rel17} (the most up-to-date 3GPP specifications are Release 18 \cite{5gsys18, 5gsec18, 5gproc18}), and it is the only one that has an OAuth 2.0 access control mechanism implemented, compared with Open5GS \cite{open5gs} and OAI \cite{oai}. 

 We only focused on the non-roaming mode scenario, so we do not consider the specific NFs and proxies that are used for roaming mode purposes. 
 We study the inter-NF interactions, not the UE-RAN-NF communication part. Further, we consider the direct NF communication mode where NFs interact directly with each other, and do not cover the indirect communication mode where the Service Communication Proxy (SCP) intermediates between NFs to handle all requests and responses. 
 
 We want to find access privilege escalation attacks or core network disruption attacks by assuming a compromised NF sending malicious messages (or an attacker obtaining core network communication access to send messages, possible in a no-TLS scenario). We choose a testing-based methodology against the most mature implementation \freefivegc, focusing on the specific perspective of core-NF interactions, instead of UE-RAN attack surfaces chosen by many previous works \cite{corecrisis, 5gfuzzinfocomm}, and we want to systematically cover a broad range of core NF API functionalities, instead of only one API of one NF as done in a previous work \cite{syedapifuzzing}.


%% file: Template/sections/Casestudy.tex
\section{\ourattack}
\label{sec:cases}
In this section, we first present a detailed threat model we assume in this paper, then describe a very severe access control vulnerability we found manually, \ourattack, serving as motivation for the design of \ourfuzz. 

\subsection{Threat model} 
\label{sec:sysThreatMod}
\textbf{Attacker objectives.}
We assume that the attacker's goal is to obtain access to services that they are not authorized to access or to impact the availability of certain services.
The attacker sits under the radar and it tries to achieve their goals while interacting with the target service through the REST API standard interface. 

\textbf{Attacker knowledge.}
We consider an attacker having information about the general SBA architecture of the 5G Core, and having access to the standard OpenAPI detailing the service model of the NFs. This information is publicly available \cite{5GC_APIs_Rel17} and easily obtainable by an adversary.

\textbf{Attacker capabilities.}
The attacker can be a compromised NF or a program able to send messages to a victim NF.
We assume that the attacker can detect 
the presence of a 5G Core network in a cloud. 
As network operators have started deploying their core network in public cloud infrastructures \cite{telegerman,boostmobile}, an attacker may easily identify the presence of a 5G Core network via traffic analysis (e.g., targeting non-encrypted control information \cite{zhang2025invade}), and successively target its NFs.

 We assume that the attacker has the capability to (1) send an arbitrary valid access token request to NRF, (2) successfully receive a valid access token, and (3) send a service request (attaching the token) with unmatched parameters with the token in the request to a victim NF. That can be possible if the attacker is a compromised NF, if the attacker can steal the identity of a valid NF to send an access token request, or if the communication between services is not running over TLS. Additionally, according to \cite{tls5g}, TLS is optional; thus, an attacker can eavesdrop and modify the communication.

 We assume that the NRF is not compromised and the attacker can not change the token itself, as the token is digitally signed by the NRF.
 
An attacker can compromise an NF by taking advantage of the 5G Core cloud deployment \cite{telegerman, boostmobile} and the potential risk of cloud container privilege escalation. An attacker may gain privileged access to NFs. For example, 
as the currently existing implementation leverages Kubernetes and Docker to implement NFs and their underlying network \cite{open5gs,free5gc,oai}, the attacker may exploit known vulnerabilities for privilege escalation and lateral movements (e.g., CVE-2024-21626, CVE-2019-5736, CVE-2023-1260), gaining access to NFs and hence interacting with other  NFs at a higher privilege level.
This provides the attacker, for instance, access to valid authorization tokens.

\subsection{\ourattack~ Description}
\begin{figure}
    \centering
    \includegraphics[width=0.95\linewidth]{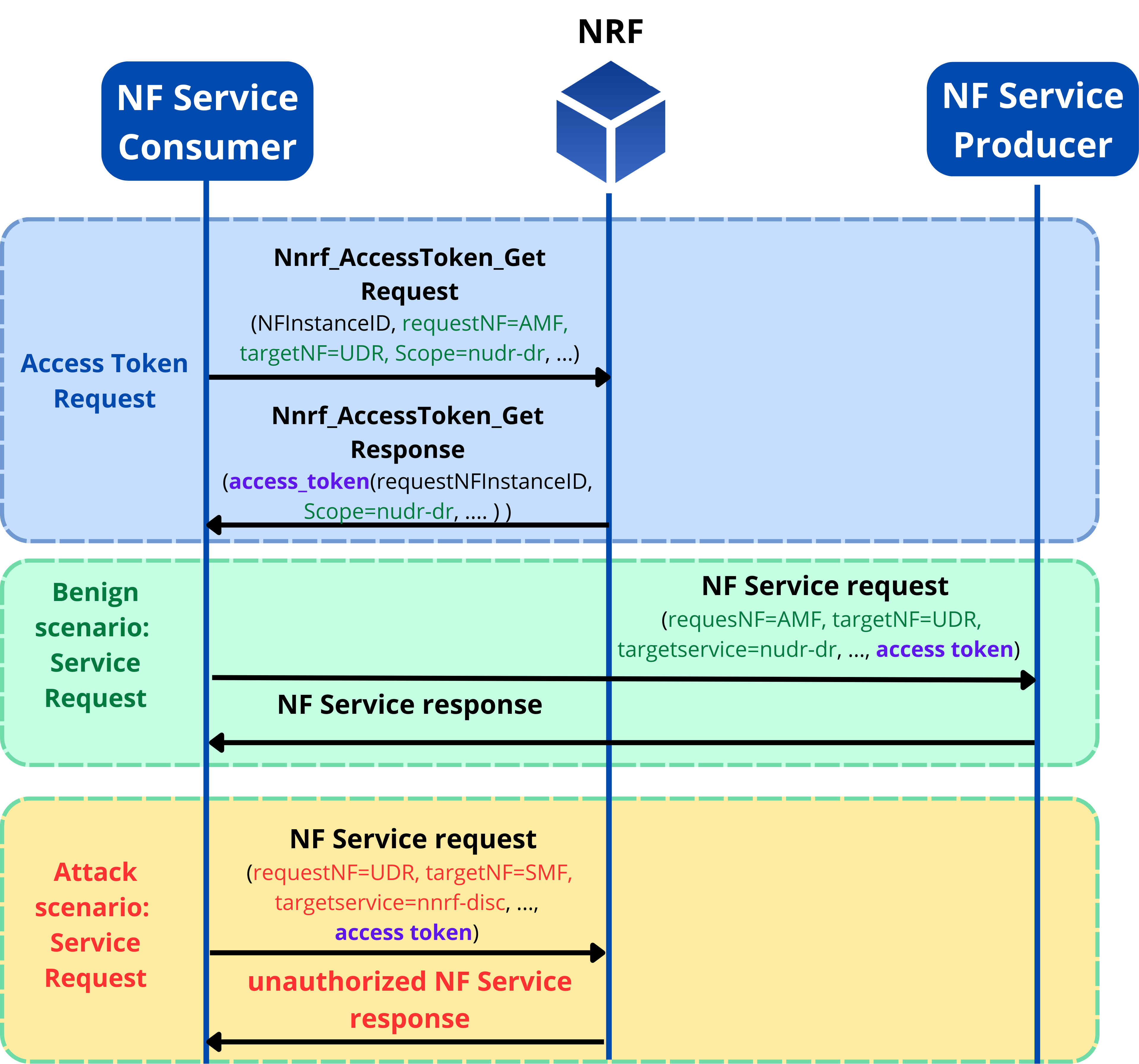}
    \caption{\ourattack}
    \label{fig:attack1}
\end{figure}

As described in the Section \ref{subsec:5gac},
when OAuth 2.0 is enabled for intercommunication between core NFs, a benign requesting NF must use a corresponding access token to visit the target NF and the target service (illustrated as the green part in Fig. \ref{fig:attack1}). However, we find that a bug in the implementation of \freefivegc{} enables the attacker to use an access token with unmatched fields for the target service (illustrated as the yellow part in Fig. \ref{fig:attack1}). 

In our experiments, we noticed that it is possible to obtain an access token from the NRF when we send an access token request with the requester NF type set to AMF, the target NF Type set to UDR, and the scope set to \texttt{nudr-dr}. The access token obtained from the NRF response message includes the scope \texttt{nudr-dr} corresponding to that of the request (illustrated as the blue part in Fig. \ref{fig:attack1}).

We then use the same access token to send a request to the NRF's \texttt{nnrf-disc} service, with the requester NF type, target NF type, and target service scope not matching those inside the token, e.g., specifying UDR as the requester NF type, SMF as the target NF type, and the target service \texttt{nnrf-disc}. Despite the mismatch between the token's scope and the requested service, the request is successfully accepted, and we get the detailed information about SMF in the response message (illustrated as the yellow part in Fig. \ref{fig:attack1}). 

This shows that the access token validation fails to enforce proper scope checks, thereby allowing unauthorized NF operations. We looked into the problem and identified that the root cause is a bug in the \freefivegc{} source code related to error handling in the \texttt{VerifyOAuth} function. 
The relevant code snippet is in \texttt{openapi/oauth/oauth.go} (Fig. \ref{lst:verify-oauth}). 
The issue occurs in the following line: \texttt{return errors.Wrapf(err, "verify OAuth scope")}.  
In this context, \texttt{err} is the result of \texttt{jwt.ParseWithClaims}, not the \texttt{verifyScope} check. 
Therefore, if \texttt{verifyScope} returns false, indicating that the access token does not have the required scope, while assuming the JWT parsing succeeds, then the \texttt{err} variable will still be \texttt{nil}. This leads to a situation where the function incorrectly returns \texttt{nil}, silently ignoring the failed scope verification. 

This bug results from forgetting to define a new error variable and assign the closest function's return value to the new error variable. Ensuring the error variable always keeps track of the latest function return value to avoid any unhandled errors is a common and famous defensive coding routine in network-related programming in the Go language.

\begin{figure}[htbp]

\begin{minted}[fontsize=\footnotesize]{go}
func VerifyOAuth(
  authorization, serviceName, certPath string,
) error {
  ...
  token, err := jwt.ParseWithClaims(...)
  if err != nil {
    return errors.Wrapf(err, "verify OAuth parse")
  }
  if !verifyScope(token.Claims.(
    *models.AccessTokenClaims).Scope, serviceName) {
    return errors.Wrapf(err, "verify OAuth scope")
  }
  return nil
}
\end{minted}

\caption{OAuth Incorrect Verification from \freefivegc }\label{lst:verify-oauth}
\end{figure}

\textbf{Impact:} The impact of this bug is broad and severe, since all the NFs rely on this OpenAPI library code for the OAuth 2.0 access control functionalities, specifically the access token scope verification. Due to the buggy logic, an access token with an arbitrary scope field value will be accepted by an arbitrary NF since that check will always pass. This could enable a huge number of possible scenarios of access privilege escalation and potential private data leakage.

%% file: Template/sections/implementation.tex
\section{\ourfuzz~System Architecture}
\label{sec:}
In this section we first describe the design goals of \ourfuzz. Then, we provide a detailed overview of its architecture.

\begin{figure*}[t]
    \centering
    \includegraphics[width=\textwidth]{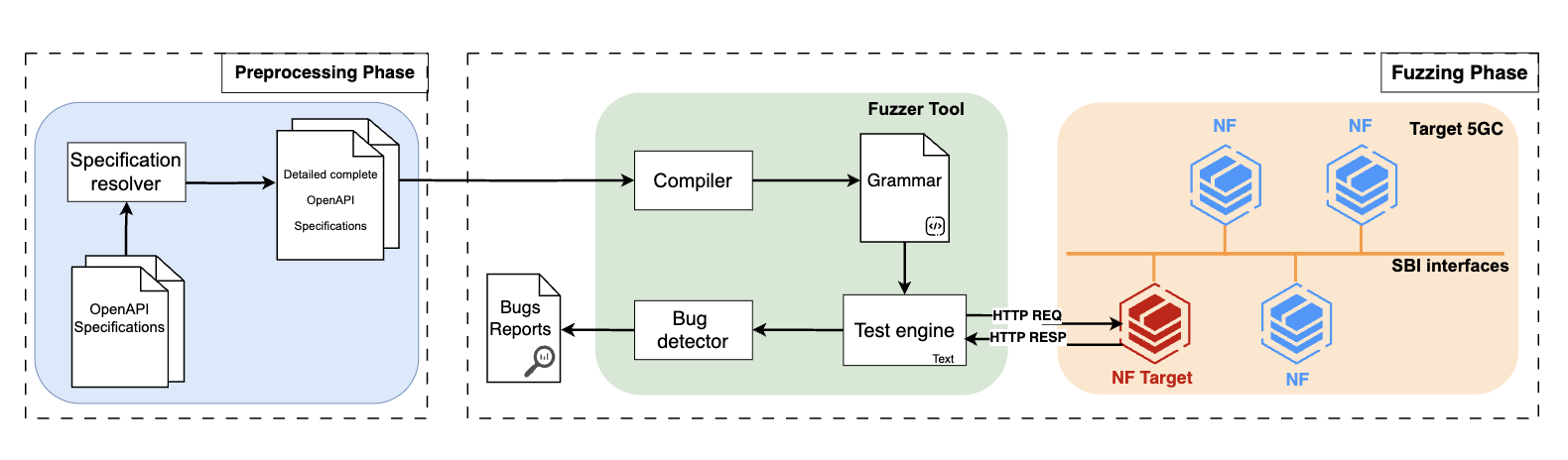}
    \caption{Overview of \ourfuzz{}}
    \label{fig:framework}
\end{figure*}

\subsection{Design Goals and Overview}
\ourfuzz{} aims to test 5G Core  open-sourced implementations
that follow the 3GPP Release 17 \cite{5GC_APIs_Rel17} specifications. We want our testing platform to achieve several goals:

\begin{itemize}
    \item \textbf{Grammar-Based Fuzzing}. All 5G Core APIs are defined by 3GPP standards, which follow a uniform RESTful framework detailing URIs, responses, and error codes. Our fuzzer should hence start from these specifications to create valid request sequences and derive useful information out of the received responses.
    \item \textbf{Stateful Approach}. Despite leveraging a RESTful framework, all 5G Core procedures are stateful. This is ensured by including all state-related information in requests, without the need for NFs to store user session information. Our fuzzer should account for the stateful nature of the system to create long testing sequences.
    \item \textbf{Protocol-Informed}. Requests among 5G Core NFs leverage secure mechanisms for authorization, such as OAuth 2.0. This means that requests are usually made in conjunction with authorization tokens, which should have a valid structure and provide at least partially correct information. Our fuzzer should account for authentication tokens to identify the fuzzable fields, while leaving unchanged essential authorization information.
\end{itemize}

Our envisioned stateful approach is to deliberately guide NF into specific and operationally meaningful states before introducing malformed or unexpected inputs. 
This broader notion of statefulness enables the evaluation of robustness and security properties under realistic runtime conditions, where specific classes of vulnerabilities are more likely to surface.

Some existing fuzzers 
already incorporate a limited form of state-awareness by learning producer–consumer dependencies between API operations described in the specification. 
For example, they may determine that a POST /sm-context request must precede a PUT /sm-context/{id} request, ensuring that the referenced resource exists before it is modified. 
This allows them to construct syntactically valid sequences that can bring a single network function into certain predictable states. In our framework, such capabilities are used as a foundation for the broader, cross-NF approach we aim to develop.

For the current experimental campaign, we adopted a configuration using RESTful API fuzzers. 
This provided wide coverage of the target SBI and allowed us to validate the framework’s end-to-end workflow, while laying the groundwork for future extensions towards fine-grained, cross-NF state guide.

\subsection{Details of \ourfuzz}

We designed a black-box testing framework to systematically assess the robustness of 5G Core SBI APIs against inputs that are either malformed, unexpected, or syntactically valid but semantically incomplete or incorrect. The framework integrates:
a target 5G Core deployment supporting OAuth 2.0 as an authorization framework, an automated fuzzing pipeline capable of executing stateful workflows derived from OpenAPI specifications, and pre-processing components to normalize and adapt 3GPP specifications for fuzzing tools.

\textbf{Specification Collection and Pre-processing}.
The goal of this phase is to assemble and prepare the OpenAPI specifications required for automated fuzzing. We began by retrieving the official 3GPP Release 17 OpenAPI specifications for the SBI from the TS 29.x series \cite{release17, 5GC_APIs_Rel17}. Each specification was then bundled \cite{redocly} into a complete, self-contained document by resolving references and merging related definitions. As part of this process, the specifications were adapted to reflect our deployment environment, including updating network-related references so that hostnames and ports matched the real topology. The resulting specifications, aligned with the actual testbed configuration, were stored in a dedicated directory and used as the input for the fuzzer.

We show an example of how we use the different specifications to resolve dependencies. In Fig. \ref{lst:udm-yaml}, we
show a fragment from the original  \texttt{TS29503\_Nudm\_SDM.yaml} (the complete specification has around 4,000 lines of YAML code). As seen in Fig. \ref{lst:udm-yaml}, the \texttt{url} is incomplete since the actual server IP address is not filled, and the \texttt{schema} field refers to another YAML specification file. After the preprocessing, however, as shown in Fig. \ref{lst:udm-bug1-yaml}, the \texttt{url} address is filled with the IP of the UDM container used in our experiment setup, and the \texttt{schema} field is resolved by referring to the content in the self YAML file, since the relevant information from \texttt{TS29571\_CommonData.yaml} has been merged into the self specification. The relevant resolved YAML specification, after pre-processing, is shown in Fig. \ref{lst:udm-bug1-yaml}.

\begin{figure}[h]
\begin{minted}[fontsize=\footnotesize]{yaml}
...
servers:
  - url: '{apiRoot}/nudm-sdm/v2'
    variables:
      apiRoot:
        default: https://example.com
        description: >
          apiRoot as defined in clause 4.4 
          of 3GPP TS 29.501.
...
paths:
  /shared-data:
    get:
      parameters:
        - ...
        -  name: supported-features
           in: query
           description: Supported Features
           schema:
             $ref: >-
                TS29571_CommonData.yaml#
                /components/schemas/SupportedFeature
        - ...
\end{minted}
\caption{Nudm\_SDM YAML specification \cite{udm-sdm-api} (partial, before resolving) }\label{lst:udm-yaml}
\end{figure}

\textbf{Compilation and Grammar Generation}.  
The pre-processed specifications is not compatible with the way RESTful fuzzers expect them. In the next step, we process these YAML specifications to create a grammar containing the fields that are the goal of our fuzzing campaign. The compilation stage produces a grammar that defines how requests can be combined into valid sequences, enabling the exploration of deeper API states.
The process take into account that the pre-processed specification follows an OpenAPI schema to infer producer–consumer dependencies between requests (e.g., ensuring a resource created by \texttt{POST} exists before being accessed by \texttt{GET} or \texttt{DELETE}).

\textbf{Test Execution}.
We use RESTler \cite{restlerpaper, restler-code} to drive our testing.
RESTler executes both individual API requests and multi-step sequences against the target SBI, instantiating parameter values using type-consistent replacements selected from fuzzing dictionaries. These dictionaries can be user-defined or automatically generated from the API specification, and constrain substitutions to values matching the expected data type.  
In addition to dictionary-based substitutions, RESTler employs a set of built-in checkers that run automatically during fuzzing to uncover issues not necessarily exposed by the main test generation logic. For example, the PayloadBodyChecker mutates request bodies by applying schema-based fuzzing rules that alter their structure, types, or values, such as dropping or duplicating fields, reordering elements, or introducing type mismatches.  
Through these mechanisms, the generated inputs range from fully valid requests with semantically varied values, to malformed or incomplete payloads that intentionally violate the API schema.  
During execution, RESTler applies a feedback-driven exploration strategy: sequences that lead to successful interactions (e.g., valid 2XX responses) are retained as building blocks for subsequent tests, pruning invalid requests from the search space and enabling deeper code exploration.

\textbf{Bug Detection and Reporting}.
When a potential fault is detected, RESTler automatically generates a bug report. By default, any response with an HTTP status code 500 (Internal Server Error) is treated as an indication of an unhandled exception and reported automatically. Additional bug reports originate from the checkers, which may target either further instances of 500 errors or specific logic flaws (e.g., hierarchy violations, resource leaks).

\textbf{Validation and Root‑Cause Analysis.}
All reported issues then undergo a manual validation step. This involves reproducing the fault, confirming that it represents an actual bug, and tracing it back to its root cause in the source code.

%% file: Template/sections/evaluation.tex
\section{Evaluation}
\label{sec:eval}

In this section, we present our experimental results. We first describe our methodology and setup, and then describe the bugs we discovered using \ourfuzz{} on \freefivegc.

\subsection{Experimental Setup}
Our evaluation was conducted on a server with 32 virtual CPUs and 64 GB of RAM, running on Ubuntu 22.04 LTS  on an x86\_64 architecture.

\textbf{\freefivegc{} setup.}
We deployed \freefivegc{} \cite{free5gc} version 4.0 with the help of the free5gc-compose \cite{free5gc-compose} repository, a Docker Compose–based distribution. The stack includes all major 5G Core NFs: AMF, AUSF,  CHF, NEF, NRF, NSSF, PCF, SMF, UDM, and UDR, as well as 3 boundary or user-plane NFs (UPF, N3iwue, and N3iwf)--- the latter ones not being our focus.
For our experiments, we removed the TNGF from the Docker Compose file, since TNGF is an access gateway function rather than a core NF, handles trusted non-3GPP access, and does not participate in the richer service interactions among core NFs. We configured the UPF to load the gtp5g \cite{gtp5g} kernel module at startup. The environment uses MongoDB 4.4 \cite{mongo} and a private Docker network for NF interconnectivity. The deployment was managed with Docker Engine v27.5.1 and Docker Compose v2.36.2.


\textbf{Specification Corpus Preparation.}
We curated and bundled a set of 48 OpenAPI specifications, drawn from the 3GPP TS 29.x (Release 17) corpus \cite{5GC_APIs_Rel17}, covering major SBI services across 10 core NFs (AMF, AUSF, CHF, NEF, NRF, NSSF, PCF, SMF, UDM, UDR). All 48 specifications were successfully compiled into RESTler grammars and used to drive the fuzzing campaign.

\textbf{RESTler setup.}
We used RESTler \cite{restler-code} version 9.3.1, deployed as a Docker container and connected to the \freefivegc{} private network for direct NF interaction. It was configured to use access-token-based authentication in Location mode, reading a pre-generated access token from a file. 
The first bug found with this setup was \ourattack~described in Section~\ref{sec:cases}.
We then leveraged this bug to test all targeted SBI services without going through the NRF, since the bug causes tokens to be accepted regardless of their intended scope. 

\subsection{Vulnerabilities Found}
Our fuzzing campaign uncovered 7 previously unknown vulnerabilities, falling into two main categories: (1) unhandled exceptions leading to runtime panics, and (2) incorrect status-code mappings. While tests included both syntactically malformed and syntactically correct inputs, the majority of these vulnerabilities were triggered by syntactically valid inputs, exploiting implementation flaws such as improper validation of optional parameters and unsafe type casting. These vulnerabilities were distributed across multiple NF within the \freefivegc{} implementation.
While we primarily worked with \freefivegc{} v4.0.0, the bugs were confirmed to be present also in \freefivegc{} v4.0.1. 
Additionally, \ourfuzz~ also rediscovered \ourattack~ we initially found manually and described in Section \ref{sec:cases}.

\subsection{Results: Runtime Panics}
Our fuzzing campaign revealed several vulnerabilities that cause runtime panics in the affected NFs. These failures, often originating from unhandled exceptions or unsafe type operations, terminate request processing and result in a 500 Internal Server Error response.

\noindent\textbf{(1) UDM — Nudm\_SDM Shared Data Retrieval: omission of \texttt{supported-features} causes panic.}  
The Nudm\_SDM service is responsible for providing UE subscription data to other NFs via the Nudm SBI interface. In the Shared Data Retrieval operation (\texttt{GET /nudm-sdm/v2/shared-data}), the \texttt{supported-features} query parameter is defined in the specification (Fig. \ref{lst:udm-bug1-yaml}) as optional (since \texttt{required: true} is not specified, and OpenAPI v3.0 treats all request parameters as optional by default \cite{swagger-openapi}) and intended for feature negotiation between client and server.  

We found that omitting this parameter in an otherwise syntactically valid request causes the UDM to terminate with a runtime panic, returning a \texttt{500 Internal Server Error}. Such a condition can be triggered repeatedly to cause NF-level denial of service.

\begin{figure}[htbp]
\begin{minted}[fontsize=\footnotesize]{yaml}
...
servers:
- url: http://udm:8000/nudm-sdm/v2
...
paths:
  /shared-data:
    get:
        ...
        parameters:
        - ...
        - name: supported-features
          in: query
          description: Supported Features
          #required: false
          schema:
            $ref: >-
                #/components/schemas/
                SupportedFeatures
        - ...
\end{minted}
\caption{Nudm\_SDM YAML specification (partial, after resolving) }\label{lst:udm-bug1-yaml}
\end{figure}

Our code inspection revealed that the request is processed by the \texttt{HandleGetSharedData} function (Fig. \ref{lst:udm-bug1}), which retrieves the parameter using \texttt{c.QueryArray("supported-features")} and immediately accesses the first element without verifying the array length. When the parameter is absent, this leads to an out-of-bounds access and an unhandled exception.  

This behavior violates the API specification (Fig. \ref{lst:udm-bug1-yaml}), which allows the parameter to be omitted, and highlights the lack of proper input validation for optional parameters in the UDM implementation.

\begin{figure}[htbp]
\begin{minted}[fontsize=\footnotesize]{go}
// GetSharedData - retrieve shared data
func (s *Server) HandleGetSharedData(c *gin.Context) 
{
   logger.SdmLog.Infof("Handle GetSharedData")
   sharedDataIds := c.QueryArray("shared-data-ids")
   supportedFeatures := c.QueryArray("supported-
   features")
   s.Processor().GetSharedDataProcedure(c, 
   sharedDataIds, supportedFeatures[0])
}
\end{minted}
\caption{Buggy code for Nudm\_SDM Shared Data Retrieval: omission of
supported-features causes panic }\label{lst:udm-bug1}
\end{figure}

\textbf{Impact:} Runtime panics in UDM translate into the unavailability of the NF, affecting its functionalities in subscriber data authentication, authorization, and mobility. Also, in the SBA, the failure of a single NF can interrupt dependent procedures, causing further service interruption. An attacker can repeatedly trigger this NF-level denial of service with minimal effort via syntactically valid inputs.

\noindent\textbf{(2) UDM — Nudm\_SDM Session Management: missing \texttt{single-nssai} causes panic.}
In the Session Management data retrieval operation (\texttt{GET /nudm-sdm/v2/\{supi\}/sm-data}), the \texttt{single-nssai} query parameter is optional and, when present, is used to filter results.

We found that omitting this parameter in an otherwise syntactically valid request causes the UDM to respond with \texttt{500 Internal Server Error}. The processing logic attempts to unmarshal the \texttt{single-nssai} value unconditionally without first checking whether the parameter is present. When it is absent (empty string), the unmarshal fails and the error propagates as a \texttt{500 Internal Server Error}. In fact, if \texttt{single-nssai} is absent, the request should be processed normally without NSSAI filtering (i.e., success \texttt{200 OK}). 

\textbf{Impact:}  This attack reveals incorrect code logic: it attempts to unmarshal a value unconditionally without first checking whether the parameter is present. Propagating internal server errors might cause management intervention efforts to investigate the error root cause, resulting in unnecessary service downtime, with the effect of DoS attacks.

\noindent\textbf{(3) UDM — Nudm\_SDM Session Management: malformed \texttt{single-nssai} causes runtime panic.}
In the Session Management data retrieval operation (\texttt{GET /nudm-sdm/v2/\{supi\}/sm-data}), the \texttt{single-nssai} query parameter is optional and, when present, is used to filter results. We found that putting a malformed value for this parameter causes the UDM to respond with \texttt{500 Internal Server Error}.  When \texttt{single-nssai} is present but malformed, the service should return a client error (\texttt{400 Bad Request}) rather than \texttt{500 Internal Server Error}.

\textbf{Impact:}  This attack reveals incorrect code logic: when faced with a malformed value, it fails to check that and classify it as client-side bad behavior but claims it as a server-side error. Propagating internal server errors might cause management intervention efforts to investigate the error root cause, resulting in unnecessary service downtime with the same effect as DoS attacks.

\noindent\textbf{(4) UDM — Nudm\_SDM Shared Data Subscription: invalid parameters cause runtime panic.}

In the Shared Data Subscription operation (\texttt{POST /nudm-sdm/v2/shared-data-subscriptions}), the request body contains parameters such as the NF instance identifier, a callback reference, and monitored resource URIs. These are defined as strings in the specification but may require additional semantic validation by the service logic.

We found that when the request body contains syntactically valid JSON but semantically invalid parameter values, the UDM responds with \texttt{500 Internal Server Error}. The processing logic deserializes the request body without enforcing strict validation of parameter correctness; as a result, invalid values trigger a failure during execution, which is then mapped to a generic server fault.

The service should detect invalid parameter values at the validation stage and return a client-side error (\texttt{400 Bad Request} or other appropriate \texttt{4xx} code) instead of \texttt{500 Internal Server Error}.

\textbf{Impact:} The code does not provide protection against malicious parameters and fails to classify the attack into client-side bad behavior. Instead, the implementation incorrectly claims it as a server-side error. Propagating internal server errors might cause management intervention efforts to investigate the error root cause, resulting in unnecessary service downtime with the same effect as DoS attacks.

\noindent\textbf{(5) NSSF — NSSAI Availability Subscription: omission of \texttt{expiry} field causes panic}

The Nnssf\_NSSAIAvailability service enables other NFs to subscribe, via the Nnssf SBI interface, to notifications about the availability of network slices, identified by their NSSAI ID. In the Subscription Creation operation (\texttt{POST /nnssf\allowbreak-nssaiavailability\allowbreak/v1\allowbreak/nssai\allowbreak-availability\allowbreak/subscriptions}), the \texttt{expiry} field is specified in the standard as optional, indicating the subscription’s expiration time.

We found that omitting this field in an otherwise syntactically valid request causes the NSSF to terminate with a runtime panic, returning a \texttt{500 Internal Server Error}.

Our code inspection revealed that the request is handled by the \texttt{NssaiAvailabilitySubscriptionCreate} function, which unmarshals the request body into a structure where \texttt{expiry} is a pointer to a \texttt{time.Time} object. The implementation immediately calls \texttt{.IsZero()} on this pointer without verifying whether it is \texttt{nil}. When the field is absent, the pointer remains \texttt{nil}, resulting in a nil pointer dereference and an unhandled exception.

This behavior violates the API specification, which allows the omission of the \texttt{expiry} field, and underscores the lack of proper input validation for optional parameters in the NSSF implementation.

\textbf{Impact:} Runtime panics in NSSF translate into the unavailability of the NF, affecting its functionalities in selecting appropriate network slices for devices based on subscription and service requirements. Also, in the SBA, the failure of a single NF can interrupt dependent procedures, causing further service interruption. An attacker can repeatedly trigger this NF-level denial of service with minimal effort via syntactically valid inputs.

\noindent\textbf{(6) PCF — BDT Policy Creation: unsafe type cast in request handling causes panic}

The Npcf\_BDTPolicyControl service allows other NFs to request data transfer policies from the PCF via the Npcf interface. In the Policy Creation operation (\texttt{POST /npcf-bdtpolicycontrol/v1/bdtpolicies}), the request body is deserialized into a \texttt{BdtReqData} structure representing the requested transfer parameters.

We found that sending a syntactically valid request triggers a runtime panic in the PCF, returning a \texttt{500 Internal Server Error}. The crash stems from an unsafe type assertion in the \texttt{HandleCreateBDTPolicyContextRequest} function, which directly casts the result of \texttt{deepcopy.Copy(requestMsg)} to \texttt{*models.BdtReqData} without checking its type.

This flaw can be exploited to repeatedly crash the PCF, resulting in NF-level denial of service, and highlights the absence of type-safety checks in the BDT policy creation logic.

\textbf{Impact:} Runtime panics in PCF translate into the unavailability of the NF, affecting its functionalities in providing policy and QoS rules to control how user traffic is handled. Also, in the SBA, the failure of a single NF can interrupt dependent procedures, causing further service interruption. An attacker can repeatedly trigger this NF-level denial of service with minimal effort via syntactically valid inputs.

\subsection{Results: Incorrect Status-Code Mapping}

Our fuzzing campaign also uncovered one case of incorrect status-code mappings, where the NF returned an HTTP status code inconsistent with the actual outcome of the operation. In this case, the error condition was detected but associated with the wrong status code, hindering proper error handling.

\noindent\textbf{(7) UDM — Nudm\_SDM ID Translation: non-existent \texttt{ueId} yields 500 status code instead of 404}
In the ID Translation Result operation (\texttt{GET /nudm-sdm/v2/{ueId}/id-translation-result}), the request retrieves identity translation data for a given UE identifier.

We found that when the \texttt{ueId} does not exist in the UDR, the UDM responds with \texttt{500 Internal Server Error} instead of propagating the \texttt{404 Not Found} returned by the UDR. The request is forwarded to the UDR, which correctly returns a \texttt{404}. However, the UDM does not handle the UDR’s \texttt{404 Not Found} response correctly and instead returns a generic \texttt{500 Internal Server Error} with a \texttt{SYSTEM\_FAILURE} cause.

This constitutes an incorrect status-code mapping: the UDM should handle \texttt{404 Not Found} responses from the UDR explicitly and propagate the correct status code to the API consumer instead of returning a generic \texttt{500}.

\textbf{Impact:} It fails to utilize the \texttt{404 Not Found} response from UDR as the correct reply, but claims the problem as a server-side error. Propagating internal server errors might cause management intervention efforts to investigate the error root cause, resulting in unnecessary service downtime with the same effect as DoS attacks.

\subsection{Discussion}
During our evaluation we also identified several limitations of the current \freefivegc{} implementation. 

\textit{Access Token Request Implementation.}
When a customer NF contacts the NRF to request an access token to visit a producer NF' specific service (or service list), it must include its NFInstanceID, NFtype, the target network function type, target service scope (might be a list), and some optional parameters. Examples of such parameters include RequesterPlmn, RequesterPlmnList, RequesterSnssaiList, RequesterFqdn, RequesterSnpnList, TargetPlmn, TargetSnpn, TargetSnssaiList, TargetNsiList. \freefivegc{} implements the main functionality, however, at the NRF side, it does not implement the related logic to cross-check the validity of the various complex optional parameters for making the decision of whether or not to issue an access token \cite{free5gcatcode}. 

When generating an access token, \freefivegc{} current code tries to record a limited number of parameters in the access token claim data structure, namely the NRF's  NFInstanceID, the customer NF's NFInstanceID, the target NF's NFInstanceID, the service scope, and the token's expiration time \cite{free5gcatcode}. However, they have not implemented the code to record the NRF Instance ID and the target NF Instance ID, both of which are mandatory to be included in the access token content according to the specifications \cite{5gsec15, 5gsec18}. The only recorded information is the customer NF Instance ID, the service scope, and the token's expiration time. Lastly, the various complex optional parameters are not recorded in the final access token.

\textit{Access Token Validation Implementation.}
Meanwhile, the validation logic at the provider side, upon receiving an access token, is also very limited. The only validation is to see if the target service name is inside the scope list in the access token and if the scope field of the access token is not vacant \cite{free5gcatusecode}.

%% file: Template/sections/relatedWork.tex
\section{Related Work}
\label{sec:related}

\textbf{Formal Analysis.}
The work in \cite{formal5gac} formally analyzed the OAuth-based access control mechanism of the 5G Core in the direct non-roaming mode as specified in 3GPP Release 17 \cite{release17}, employed several techniques to construct the abstract model, and tested it against 55 properties. This led to the discovery of five vulnerabilities, including unauthorized access to sensitive information, illegitimate access to services, and denial of service. The work in \cite{sp25formal5g} also worked on the formal analysis of the 5G Core access control system, including the indirect communication mode and 5G roaming. It analyzed 61 security properties across six model components and revealed 10 distinct types of access control attacks, including five newly discovered attacks.

\textbf{UE and Base Station Input Fuzzing.} The work in \cite{ransacked} fuzzed cellular interfaces accessible from a base station or user device, compiled ASN.1 specifications into structure-aware fuzzing modules, performed fuzzing on seven open-source and commercial cores, revealing 119 vulnerabilities. The work in \cite{corecrisis} used automata learning methods to infer the FSM of AMF, SMF, and their interaction. It developed a positive feedback cycle of iterative testing and FSM refinement, introduced state-aware mutation techniques, and evaluated on three open source and one commercial 5GC implementation, uncovering 15 exploitable vulnerabilities that could lead to DoS, authentication bypass, and billing fraud. The work in \cite{5gfuzzinfocomm} designed a state transition feedback guided fuzzing mechanism, tested on 3 5GC implementations (Open5GS\cite{open5gs}, \freefivegc{} \cite{free5gc}, and OAI-5GC\cite{oai}), and discovered 22 real vulnerabilities. The work in \cite{amfuzz} targeted only the AMF fuzzing, since it exposed the attack surface for UE RAN inputs and also tested on the above 3 implementations.

\textbf{5G Core Network API Fuzzing.} The ongoing work in \cite{syedapifuzzing} customized the RESTler \cite{restlerpaper} tool, targeting to find flaws in 5GC inter-NF communication API implementations, and designed a state-aware feedback-driven fuzzing framework, where the authors use response codes and messages to detect issues or augment the fuzzing corpus with new messages.
To generate test inputs, the authors leverage five mutation strategies, i.e., parameter values alteration, message swap, message deletion, message addition, and message combination. It was tested on the \freefivegc{} \cite{free5gc} AMF EventExposure API, and uncovered two previously unknown vulnerabilities. 

\textbf{Specification Inconsistency.} The work in \cite{detectincocell} used a few-shot learning mechanism on domain-adapted large language models, detected inconsistencies of the Non-Access-Stratum (NAS) and the security specifications of 4G and 5G networks, uncovered 157 inconsistencies, and validated them on three open-source implementations and 17 commercial devices.

%% file: Template/sections/conclusion.tex
\section{Conclusion}
\label{sec:conclusion}
The service-based architecture of the 5G Core network provides significant advantages to operators, but also opens up novel possibilities for attackers that compromise an NF or are able to obtain a valid token. 
To demonstrate that such attacks are possible, we first crafted an attack against \freefivegc, the only open source 5G Core implementation supporting OAuth access control. 
The attack, which we call \ourattack~ allows an attacker to obtain unauthorized access to resources using a valid token issued for a service, to obtain unauthorized access to resources managed by a different service.
We then proposed \ourfuzz, a grammar-based fuzzing approach to test SBIs for unhandled error messages, misconfigurations, or authorization failures.
Testing it on \freefivegc, we identified seven previously unknown vulnerabilities in three different services, causing denial of service and improper error handling. \ourfuzz~ also automatically rediscovered our \ourattack.
All bugs were acknowledged by the developers, 7 were already patched, and the only remaining bug has a patch under development. 
Our results show that tools for the automatic identification of vulnerabilities tailored to the 5G Core are essential to secure the cellular network critical infrastructure.